# From Text to Bank Interrelation Maps

Samuel Rönnqvist and Peter Sarlin
Turku Centre for Computer Science – TUCS
Åbo Akademi University, Department of Information Technologies
Turku, Finland
{sronnqvi, psarlin}@abo.fi

***Abstract*—In the wake of the ongoing global financial crisis, interdependencies among banks have come into focus in trying to assess systemic risk. To date, such analysis has largely been based on numerical data. By contrast, this study attempts to gain further insight into bank interconnections by tapping into financial discussion. Co-mentions of bank names are turned into a network, which can be visualized and analyzed quantitatively, in order to illustrate characteristics of individual banks and the network as a whole. The approach allows for the study of temporal dynamics of the network, to highlight changing patterns of discussion that reflect real-world events, the current financial crisis in particular. For instance, it depicts how connections from distressed banks to other banks and supervisory authorities have emerged and faded over time, as well as how global shifts in network structure coincide with severe crisis episodes. The usage of textual data holds an additional advantage in the possibility of gaining a more qualitative understanding of an observed interrelation, through its context. We illustrate our approach using a case study on Finnish banks and financial institutions. The data set comprises 3.9M posts from online, financial and business-related discussion, during the years 2004 to 2012. Future research includes analyzing European news articles with a broader perspective, and a focus on improving semantic description of relations.**

## I. Introduction

The still ongoing global financial crisis has brought several banks, not to say entire banking sectors, to the verge of collapse. Considering the costs of banking crises, the recent focus of research on financial instabilities is well-motivated. First, real costs of banking crises have been estimated to average at around 20-25% of gross domestic product (GDP) (e.g., [1,2]). Second, data from the European Commission illustrate that government support for stabilizing banks in the European Union (EU) peaked at the end of 2009. The support amounted to €1.5 trl, which is more than 13% of EU GDP. The current financial crisis has stimulated a particular interest in linkages, interrelations, and interdependencies among banks, and the overall importance of systemic risk assessment and identification.

Most common sources for describing bank networks are based upon numerical data like interbank asset and liability exposures, and co-movements in market data (e.g., equity prices, CDS spreads, and bond spreads) (see [3]). While these direct and indirect linkages complement each other, they exhibit a range of limitations. First, interbank data, while measuring direct linkages between banks' balance sheets, are mostly not publicly disclosed. In many cases, even regulators have access to only partial information. Second, while being widely available and capturing other contagion channels than those in direct linkages between banks [4], market price data assume that asset prices correctly reflect all publicly available information on bank risks, exposures and interconnections. Yet, it has repeatedly been shown that securities markets are not always efficient in reflecting information about stocks (e.g. [5]). In addition, market prices are most often contemporaneous, rather than leading indicators, and it might be difficult to separate the factors driving market prices in order to observe bilateral interdependence [6].

This paper proposes an approach to assessing relationships among banks by analyzing how they are mentioned together in financial discussion, such as news, official reports, discussion forums, etc. The idea of analyzing relations in text is in itself simple, but widely applicable. It has been explored in various areas; for instance, Özgür et al. [7] study co-occurrences of person names in news, and Wren et al. [8] extract biologically relevant relations from research articles. These approaches can be used to construct social or biological networks, using text as the intermediate medium of information. Much of the work on relation extraction focuses on semantic detail rather than large scale; we choose to focus on large scale primarily. Our contribution lies in introducing this text-based approach to the study of bank interrelations, with emphasis on targeting analysis of the resulting network models.

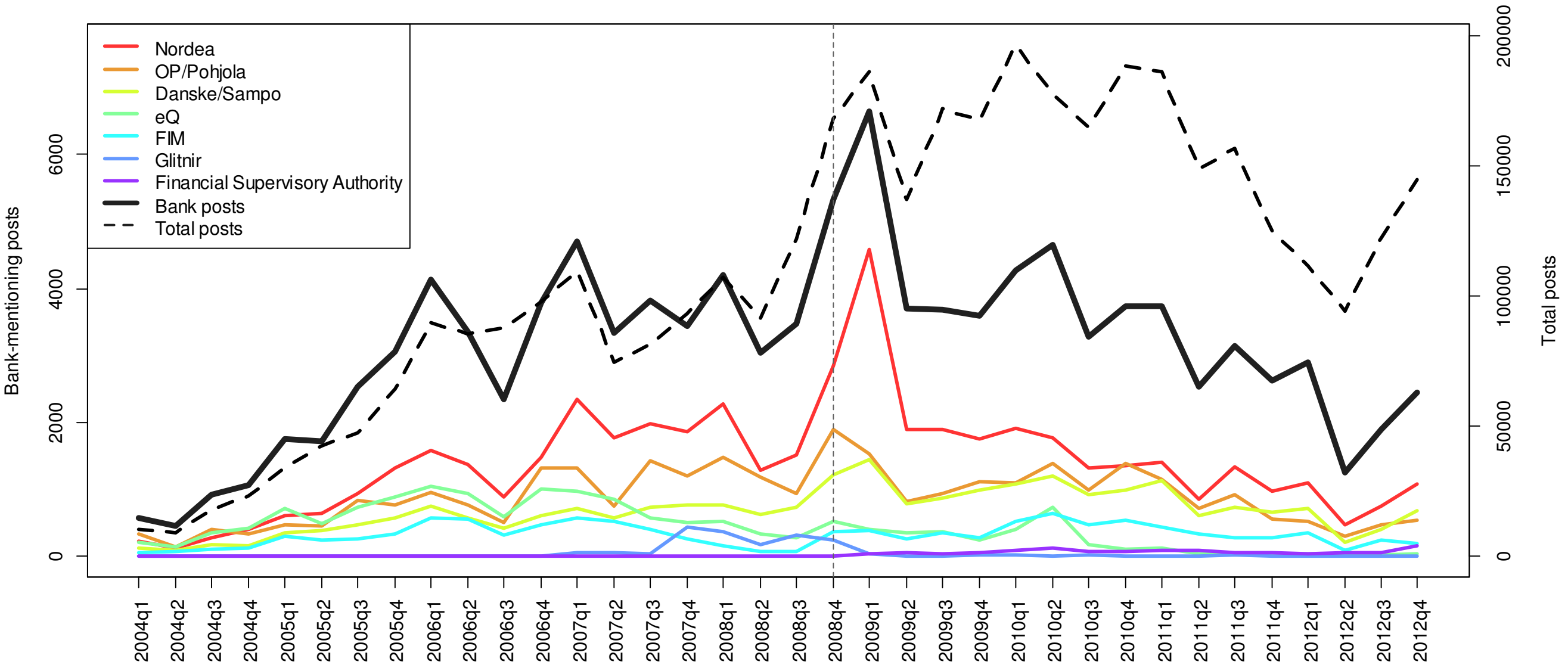

Fig. 1. Discussion volume trends of a subset of banks and total posts and total bank posts. The left scale refers to the volume of posts that mention any studied bank and the right scale to the number of posts in total (bank posts and other). The vertical line marks the quarter following the financial crash of 2008.

On the one hand, our approach may be compared to the above discussed, more established ways of quantifying bank interdependence. While not measuring direct interdependence, it has the advantage over interbank data by relying upon widely available data, and over co-movements in market data by being a more direct measure of an interrelation. Further, the co-movement-based approaches, such as that in Hautsch et al [9], require large amounts of data, often invoking reliance on historical experience, which may not represent the interrelations of today. On the other hand, our approach serves to shed light on banks' relationships in the public view, offering a perspective different to tried methods, especially considering the presence of rich, embedded contextual detail. Thus, rather than an ending point, this sets a starting point from which further analysis may focus more extensively on the context of mentions.

We exemplify our mapping approach through a case study, which focuses on Finnish banks discussed in Finland. A co-mention network is derived from 3.9M posts, covering the years 2004 to 2012, in an online discussion forum belonging to a major Finnish web portal dedicated to financial and business information. This paper discusses some observations that can be made based on a network, as well as directions in which this approach could be extended. Text as source data holds plenty of potentially interesting detail on banks and their relations, but its interpretation by computational methods is often challenging.

The co-mention network illustrates relative prominence of individual banks, and segments of more closely related banks. The dynamics of the network, both local and global, reflect real-world events over time. The network can also be utilized as an exploratory tool that provides an overview of a large set of data, while the underlying text can be retrieved for more qualitative analysis on relations. Yet, we acknowledge that the data of this case study has its limitations. Hence, it only sets a basis for a more thorough and complete assessment of bank interrelations, where the focus could be on European banks and European news articles, for instance. Additionally, we look to introduce more semantic detail to describe relations, and thus facilitate interpretation, in future work.

This paper is structured as follows. Section 2 describes the text data and methods used in this paper. Section 3 presents the derived mappings and their use for static and dynamic assessment of interrelations. Section 4 concludes and discusses future research directions.

## II. Data and Methods

This section discusses the text data and methods used in this paper, as well as summarizes with a text-to-network process.

### A. *Data*

The text data used in this study is collected from an online discussion forum, part of a major Finnish web portal dedicated to financial and business information (Kauppalehti). The analysis focuses only on banks and financial institutions operating in Finland, to avoid geographical bias. The data set contains 3.9M posts, covering the years 2004 to 2012.

We hand-picked the banks to include major and smaller ones, both consumer and investment banks. The Bank of Finland (the central bank) and the Finnish Financial Supervisory Authority were included as well, since connections to these entities, in particular the latter, may provide interesting information on banks in distress. Banks that have merged or changed name during the studied time period are treated as one entity, e.g. Danske/Sampo and OP/Pohjola. The chart in Fig. 1 provides an overview of the trends in discussion volume, both general and bank-related. Among all posts, 3% mention any of the targeted banks, on average.

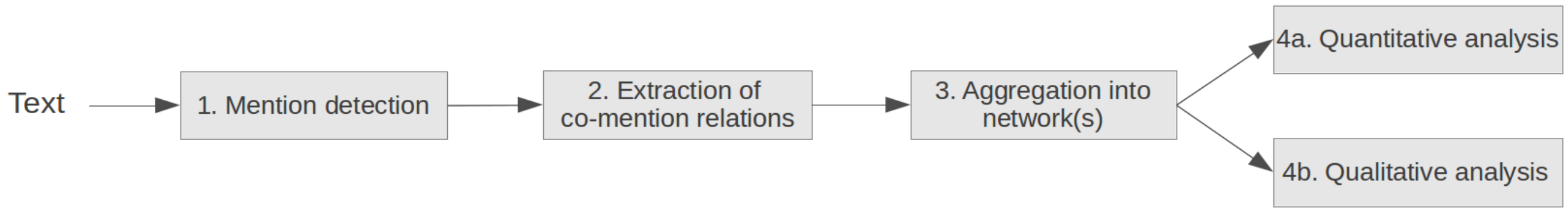

Fig. 2. Text-to-network process: (1) Mentions of bank names are detected in source text, (2) pair-wise co-mention relations are extracted between mentions within a post (context), and (3) relations aggregated over a time interval form a co-mention network. A resulting network can be analyzed with (4a) quantitative measures capturing some interesting features, and (4b) qualitative analysis through visual exploration of the network, its neighborhoods, and connectivity of individual nodes.

### B. *Methods*

Mentions and co-mentions form the basis of this study, as constituents of co-mention networks. Fig. 2 provides an overview of the process of transforming text into network models that lend themselves to analysis. With plain text as a starting point, we scan for occurrences of bank names to detect and register mentions of those banks. Scanning is performed using patterns (regular expressions) designed to match with as high accuracy as possible, accounting for word morphology of the Finnish language. A co-mention relation is formed by two bank names occurring in the same context. Multiple mentions of a single bank are counted only once per context, but a mention may participate in multiple relations. In this case, the scope of the context is defined as a single forum post. Then, a post mentioning two or more banks yields one or more pair-wise co-mention relations. We choose to disqualify any context mentioning more than 6 distinct banks, as such posts are likely to be listings or otherwise represent meaningless relations.

Aggregated into a network, the extracted relations can be studied using measures for analysis of complex networks. In the network, banks form nodes (or vertices), and aggregated co-mention relations form connections (or edges). Each connection is weighted according to the aggregated count of co-mentions, over a certain time interval. It may be desirable to enforce a threshold on weight, so that very weak connections are omitted from the network, since many existing network analysis methods are designed for binary (unweighted) networks only. Hence, unfiltered networks are more sensitive to noise, using binary measures, but low-frequency co-mentions may be of particular interest, as they are more likely to represent novel information. To illustrate the result of the text-to-network process, Fig. 3 visualizes a co-mention network based on an aggregation over the complete studied time period, 2004-2012, without weight threshold. The location of nodes in the graph visualization is based on the Fruchterman-Reingold network layout algorithm [10], so that more strongly connected parts of the network form clusters (in this case, a single center cluster).

For the analysis part, we study global network properties through quantitative measures that can indicate significant events and changes over time. We use three measures: density, average strength, and average communicability centrality. The measures are chosen to the end of illustrating a general idea: how temporal dynamics of the entire network can be quantified, in certain aspects. First, density is measured as follows:

$$D(G) = \frac{2 \cdot |E|}{|V| \cdot (|V| - 1)} \tag{1}$$

for the graph $G = (V, E)$, where $V$ is the set of vertices and $E$ the set of edges. Density is a weight-independent measure of the degree to which the network is completely connected. Second, average strength (or average weighted degree) is measured as follows:

$$\overline{S}(G) = \frac{1}{|V|} \sum_{v}^{V} S(v), \tag{2}$$

where node strength is defined as follows:

$$S(v) = \sum_{u}^{V} w_{\{u,v\}} \tag{3}$$

where $w_{\{u,v\}}$ is the weight for edge $\{u,v\}$, or 0 for non-existent edges. Third, average communicability centrality is used to measure general connectedness in the network and is measured as follows:

$$\overline{C}(G) = \frac{1}{|V|} \sum_{v}^{V} C(v), \tag{4}$$

where communicability centrality [11] (or subgraph centrality [12]) is defined as:

$$C(v) = \sum_{j=1}^{|V|} (x_j^v)^2 e^{\lambda_j} \tag{5}$$

where $x_j$ is an eigenvector of the adjacency matrix, and $\lambda_j$ is the eigenvalue. Communicability centrality reflects how densely connected the surroundings of a node is, disregarding weights. We use the average communicability centrality to describe changes in global topology, in terms of communicability centrality.

## III. Results and Analysis

This section presents and discusses the results. First, it examines the discussion volume and a co-mention network. Second, it reviews temporal changes from two perspectives: global network properties and visualized network snapshots.

### A. *Discussion volume and a co-mention network*

The data presented in Section II refer to periods before, during and after the global financial crisis, of which the collapse of Lehman Brothers may be seen as the most central event. Fig. 1 presents the trends in number of posts that mention studied banks, as well as the total post volume for reference. A notable surge in discussion activity occurred in the half year following the 2008 financial crash (2008Q4-09Q1). The second quarter saw a particular increase in mentions of Nordea, the largest consumer bank in Finland. Total post count remained high for an extended period, indicating an elevated interest in finance and business related topics in general. Overall, the larger commercial banks, Nordea, Danske/Sampo, and OP/Pohjola, are the most frequently mentioned.

The network in Fig. 2 has a dense overall topology. The strongly connected center consists of both the most popular consumer banks, such as Nordea, Danske/Sampo, and OP/Pohjola, and a number of prominent investment banks, such as eQ, Seligson, and FIM. It is also worth noting that the two supervisory authorities, the Bank of Finland and the Financial Supervisory Authority, are located in the lower right part of the network. The heterogeneity in node size (bank mention frequency), connection weight (co-mention frequency), and topology highlights the non-triviality, and need for models that describe relative bank importance and interconnection. This lends particularly for assessing potential for systemic effects of distress in individual banks.

### B. *Temporal changes: global network properties*

An aspect of central importance for interbank network models is its temporal dynamics, which can be studied both quantitatively and qualitatively. We use the introduced global network measures to profile each network instance in a single dimension, which can indicate interesting changes over time. Our selection of measures should mainly be considered a proof of concept, i.e. examples of how some specific aspects of the network can be captured. The choice of optimal measure is dependent on the eventual application, and what aspects of the network may be of interest.

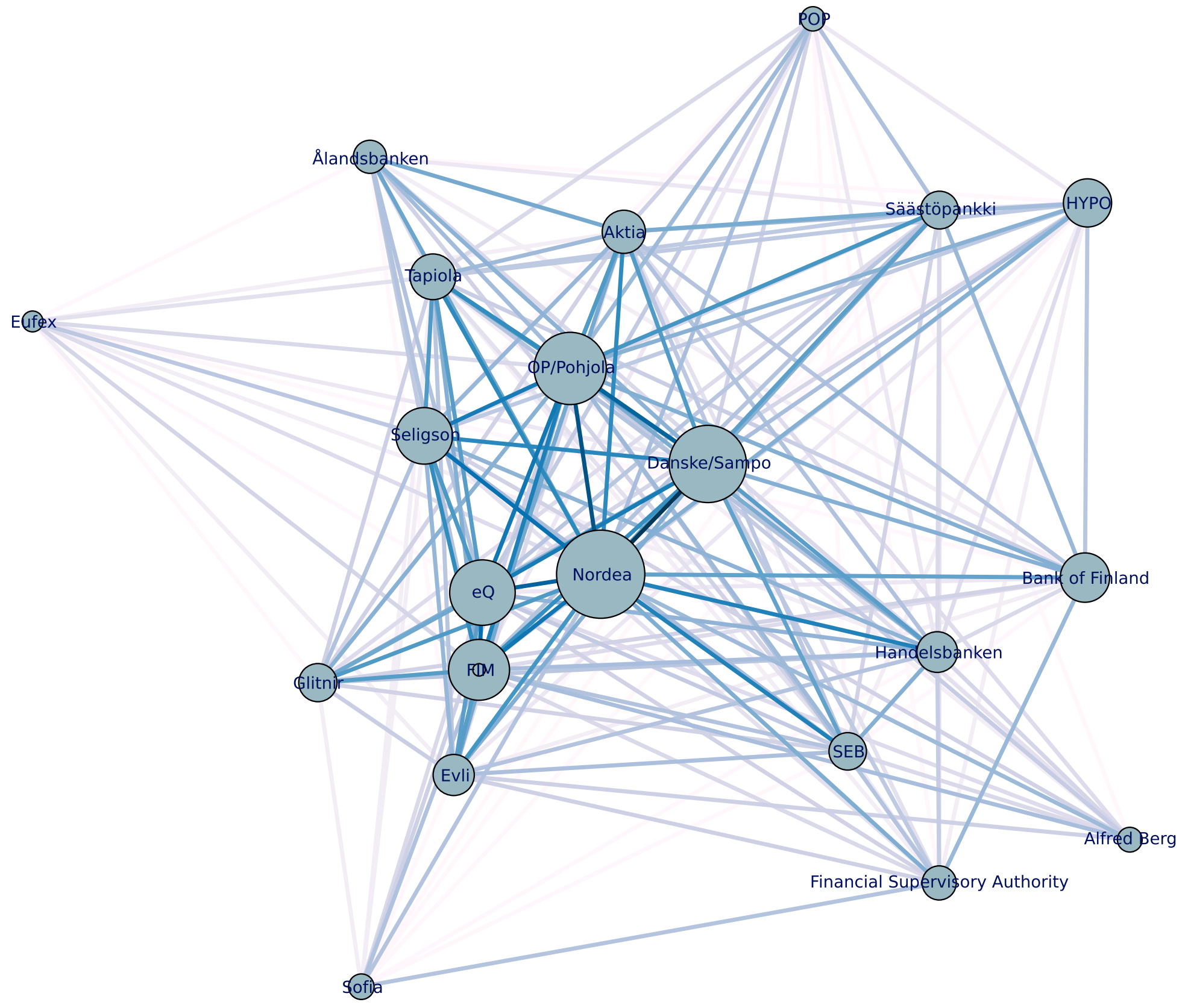

Fig. 3. Co-mention network. The network depicts counts of bank names co-occurring in forum posts for the entire period 2004-2012. Node size is proportional to the individual bank mention count, while connection darkness is logarithmically scaled to co-mention count. Nodes are positioned by the Fruchterman-Reingold algorithm [10].

The noted surge in individual bank mentions, following the financial crash of 2008, is also reflected in connection density and strength of the network, shown in Fig. 4. An even clearer reaction is visible in network communicability, in Fig. 5. A second peak occurs around 2010Q1, primarily in communicability, which might be indirectly linked to revelations concerning the financial situation in Greece and its actively-debated first bailout.

The relatively similar behavior of density and average strength hint that connection weight might not be all that vital in explaining event-related changes in discussion, rather, changing co-mention patterns may be of greater importance. Network communicability captures topological changes better than density, which does not consider the irregular structure of complex networks. It exhibits interesting event-related peaks, indicating some considerable shifts in discussion patterns in relation to eventful crisis periods.

The quantitative analysis could be focused at individual banks or groups of banks as well, to study their relations to the surrounding environment, and their centrality in the network. Given sufficient data, the information embedded in the co-mention network can also be used to generate a directed network model representing conditional probabilities of mentions, i.e. the probability of Bank B being mentioned given a context mentioning Bank A (an A→B relation). Such links could provide the basis for node centrality measures of total inbound, received attention (comparative attention on B), as

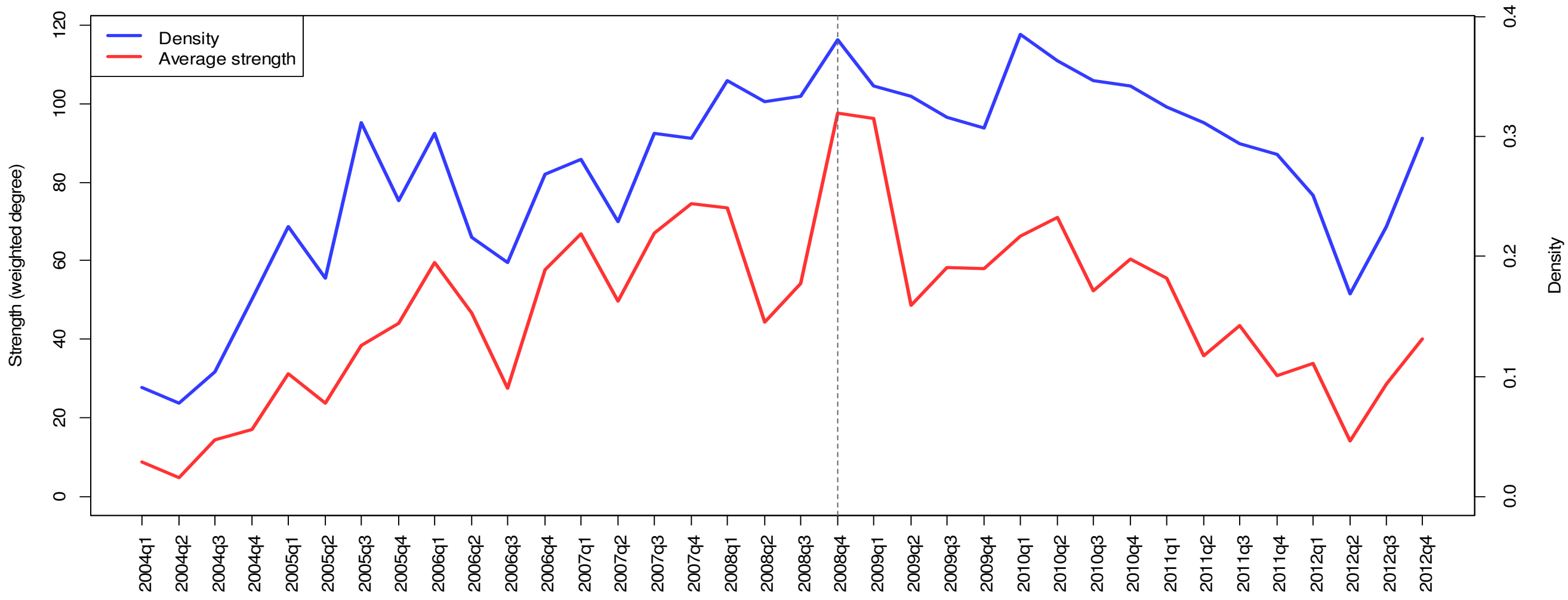

Fig. 4: Simple measures of global connectivity changing over time: average strength (left scale) and density (right scale). The vertical line marks the quarter following the financial crash of 2008.

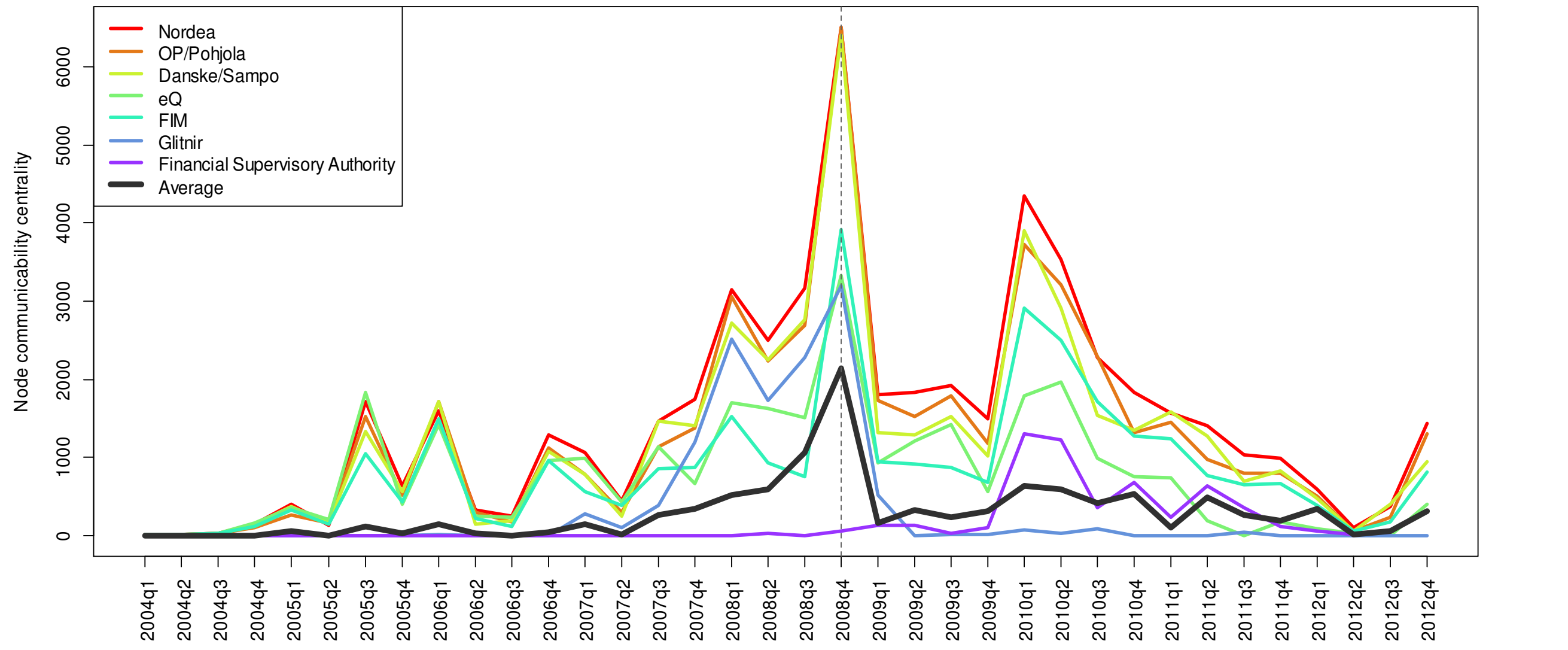

Fig. 5: Advanced measure of connectivity, communicability centrality, for a subset of relevant banks and all studied banks on average. The vertical line marks

well as total outbound, contributed attention (comparative attentiveness of A).

## C. *Temporal changes: network snapshots*

Snapshots of the network in specific time periods can be studied to understand the network, and especially parts of it, in greater detail. This section provides some brief examples of how networks can be interpreted visually, in contrast to the previous quantitative profiling. The years 2008, 2009, and 2010 represent interesting episodes, both in terms of real-world events and its reflection in network communicability. Fig. 6-8 show three snapshots of the network during each of these years, allowing for a subjective but more nuanced analysis.

In 2008, the network is highly connected, the bank Glitnir is well connected with the core of the network, and relatively many peripheral banks are co-mentioned. In 2009, the network is more concentrated to its core; peripheral banks are connected to the core in a more star-like topology. Glitnir is barely mentioned anymore, following its nationalization in 2008. The Financial Supervisory Authority is slowly emerging in comparative discussions, a sign of the changing nature of discourse following the outbreak of the crisis. In 2010, the network is again seeing increased connectivity among peripherals, especially between the Financial Supervisory Authority and the failing bank Sofia.

# IV. Conclusions and Future Research Directions

This paper has demonstrated the use of computational analysis of financial discussion, as a source for information on bank interrelations. The approach may serve as a complement to more established ways of quantifying connectedness and dependence among banks. We have shown that bank-related events, the recent global financial crisis most notably, are reflected in online financial discussion in terms of changes in bank co-mention patterns. The network of co-mentions exhibit both local and global shifts in regards to these events. However, the limitations of the current network and the underlying data occasionally lead to hazy patterns difficult to interpret and draw clear conclusions from. We suggest a number of ways these issues could be addressed in future research.

First, the scarcity in interesting events involving Finnish banks makes validation hard, in this case study. Therefore, we plan to extend this study to focus on interrelations among European banks, where more severe real-world events, such as direct failures, government interventions and distressed mergers, have occurred. In an international setting, non-apparent, cross-border bank associations may be of particular interest, as well.

Second, the nature of the co-mention relation is very wage, due to the broadness of the source discussion, and due to the relation analysis disregarding any auxiliary information that may describe the relationship in more detail. The source of data could be selected more carefully, such that the frame of discourse is narrowed down. For instance, news media articles are expected to contain a more clearly defined type of discourse, compared to an online discussion forum. The relations extracted could be further narrowed, or described, by

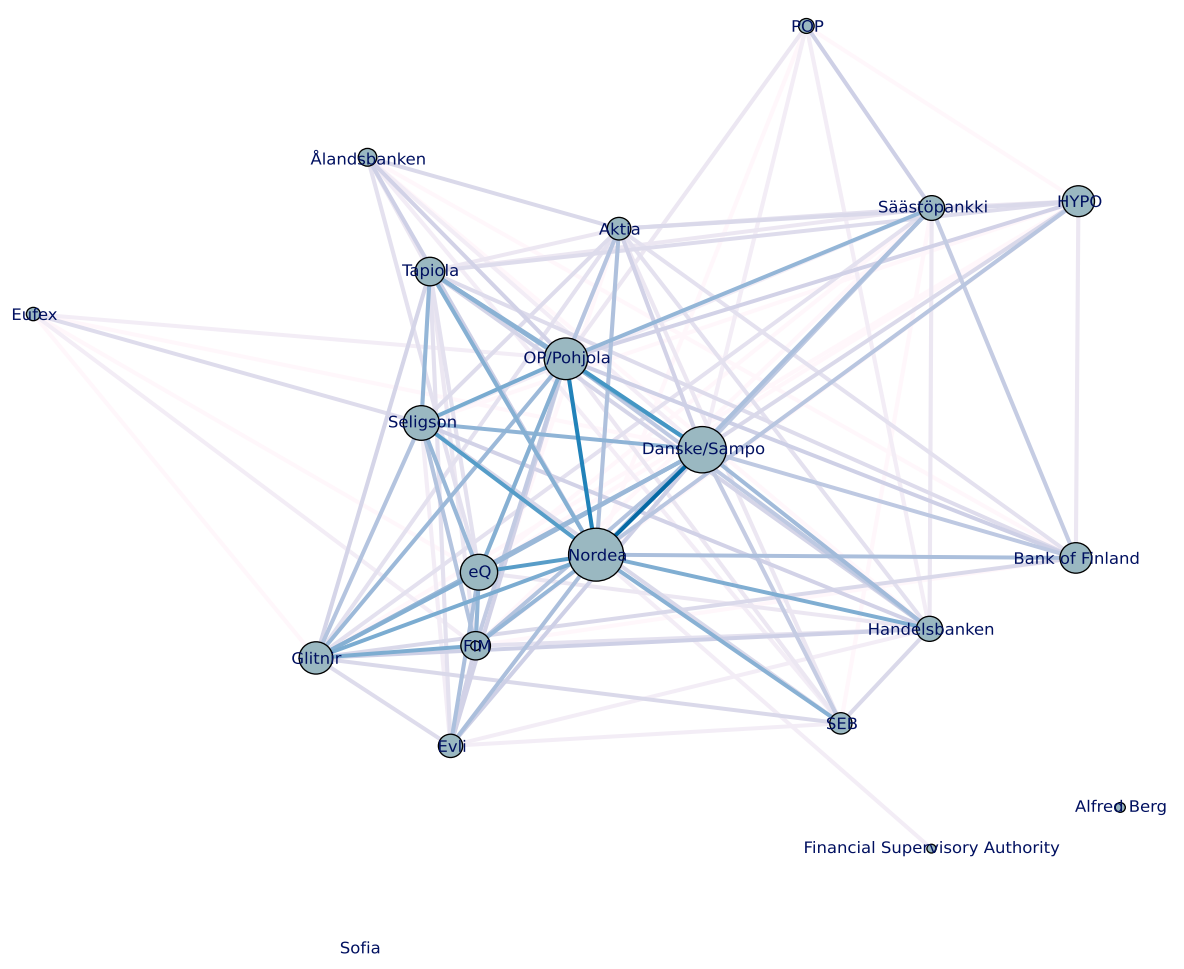

Fig. 6. A snapshot of the co-mentions in year 2008.

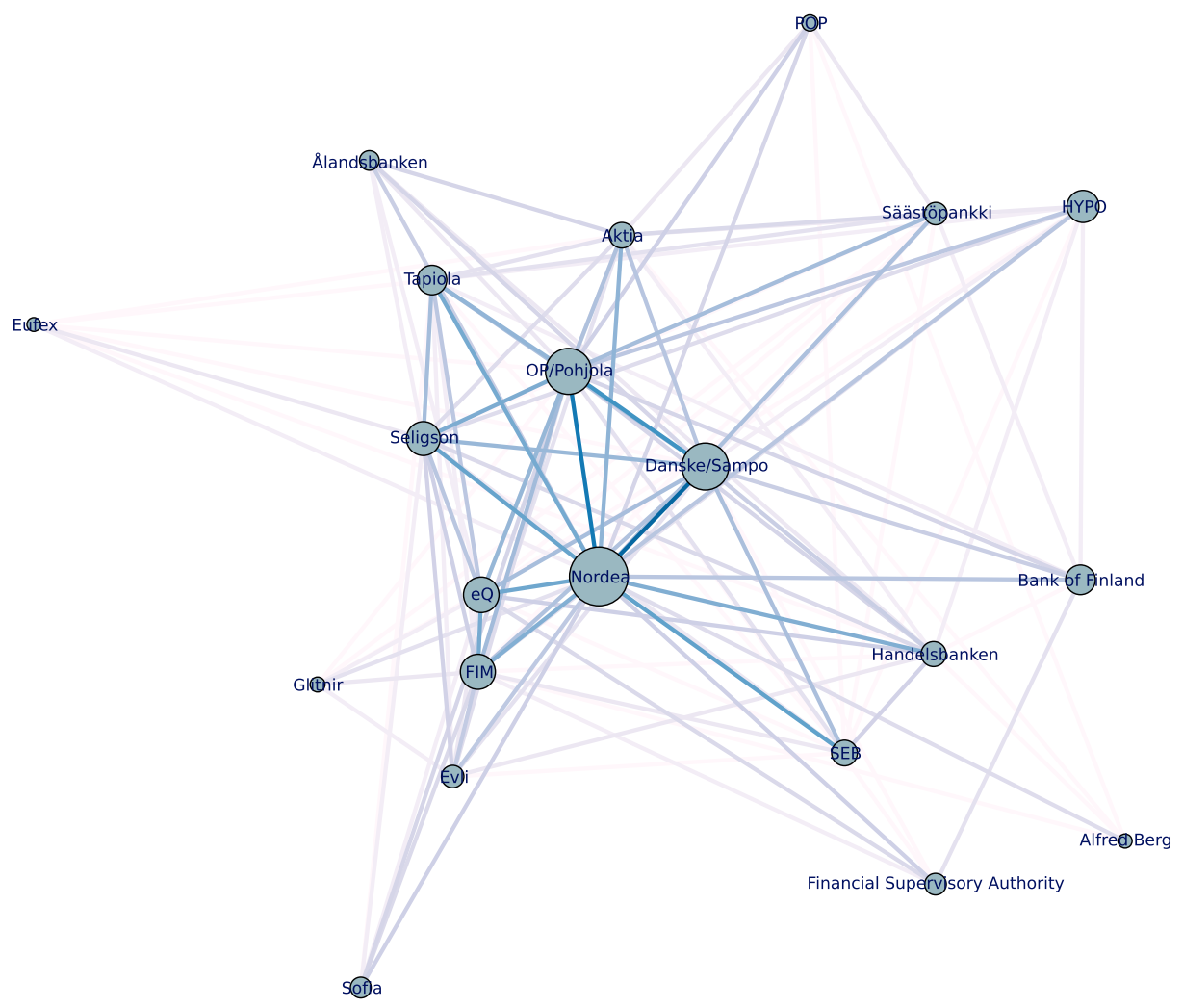

Fig. 6. A snapshot of the co-mentions in year 2009.

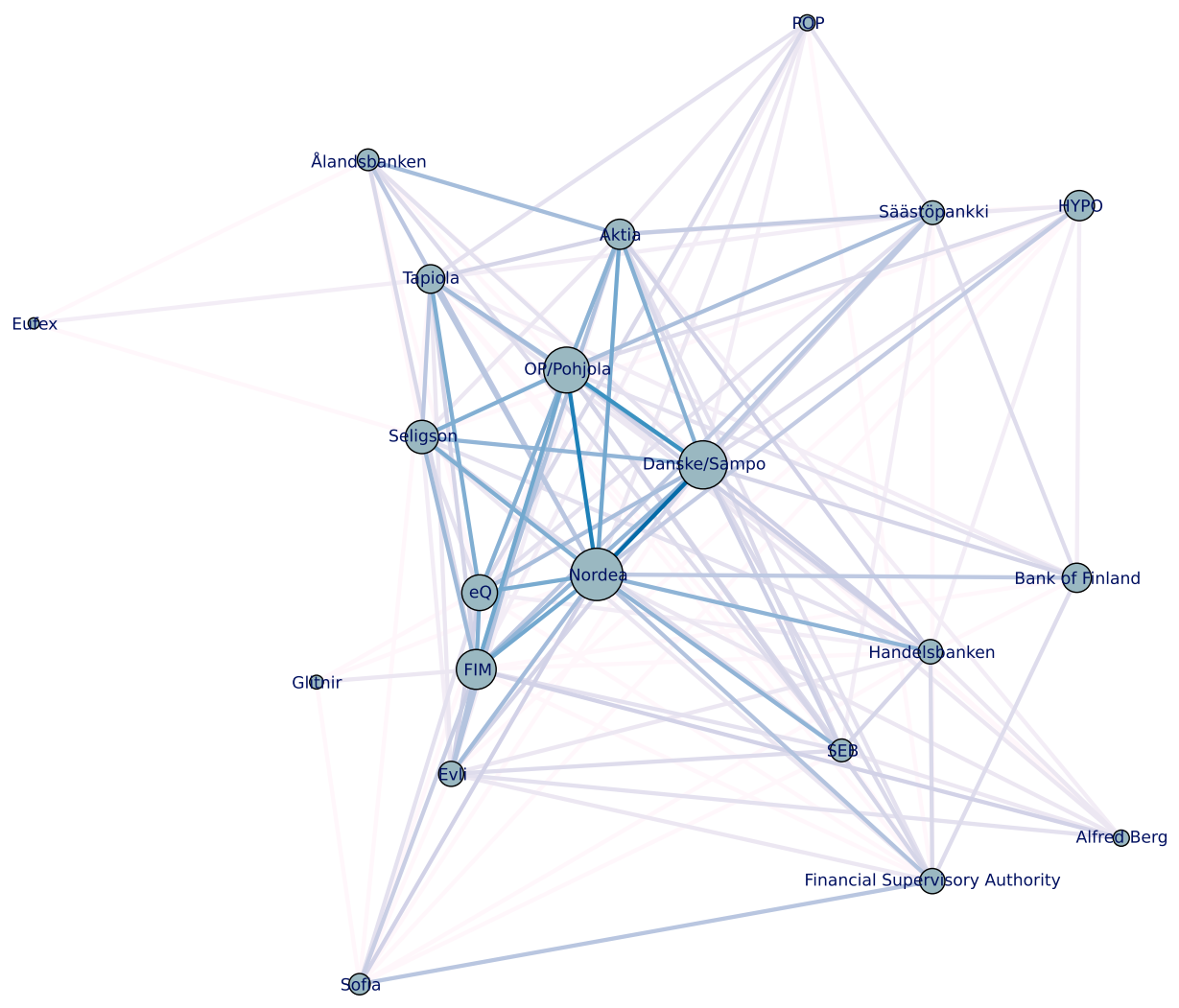

Fig. 7. A snapshot of the co-mentions in year 2010.

keywords derived from the context banks are co-mentioned in. Such an approach could in extension allow for semantic annotation of connections in the network, with potential to improve interpretability substantially.

Third, the loose definition of co-mention context, as an entire post, lessens reliability of relations extracted, while yielding more relations. Provided enough source data, the context could be narrowed down, to increase the likelihood that a relation pair is actually meaningfully related. This could be done either by shortening the context scope, e.g. to a single paragraph or sentence, or by accounting for grammatical structure (see, e.g., [13]).

Finally, methods for analyzing co-mention networks constitute a major area to work on in future research. Given a network, meaningful measures need to be applied to capture aspects of interest, be it topological or magnitudal, global or local. Still, the interpretability of such measures is fundamentally dependent on the interpretability of the network. The clearer the semantics of a relation is, the easier it is to infer intuition of any measure applied on top. This considered, our approach seems especially promising for well-enough defined targets, where meaningful and clear interpretation is possible.

## ACKNOWLEDGMENT

This paper is an extension of a poster by the authors, Tomas Eklund and Barbro Back presented at the 11th International Symposium on Intelligent Data Analysis (IDA'12) on 25–27 October 2012 in Helsinki, Finland. The authors gratefully acknowledge Academy of Finland (grant nos. 127656 and 127592) and the Graduate School at Åbo Akademi University for financial support.